\documentclass{article}
\usepackage{a4}
\usepackage{amsmath}
\usepackage[]{graphicx}
\usepackage[]{epsfig}

\newcommand{\vx}{v_x}
\newcommand{\vy}{v_y}
\newcommand{\ie}{{\em i.e.}}
\newcommand{\eg}{{\em e.g.}}
\newcommand{\etal}{{\em et al.}}
\newcommand{\dfd}{{\rm d}}
\newcommand{\vece}{\mbox{{\boldmath$\hat{e}$}}}
\newcommand{\vecv}{\mbox{\boldmath$v$}}
\DeclareMathOperator{\arctanh}{arctanh}

\begin{document}
\title{The comfortable roller coaster -- on the shape of tracks with constant normal force}

\author{Arne B. Nordmark and Hanno Ess\'en\thanks{Department of Mechanics, KTH, 100 44 Stockholm, Sweden}}
\date{July 8, 2010}
\maketitle

\begin{abstract}
A particle that moves along a smooth track in a vertical plane is influenced by two forces: gravity and normal force. The force experienced by roller coaster riders is the normal force, so a natural question to ask is: what shape of the track gives a normal force of constant magnitude? Here we solve this problem. It turns out that the solution is related to the Kepler problem; the trajectories in velocity space are conic sections.
\end{abstract}

\section{Introduction}
Motion along curved trajectories is always associated with a normal force which according to kinematics is given by,
\begin{equation}\label{eq.normal.force.kinematic}
F_n = m \frac{v^2}{\rho},
\end{equation}
where $v$ is the speed and $\rho$ the radius of curvature of the track. If the only physical force is the constraint (or reaction) force from the track, constant normal force simply requires $v \propto \sqrt{\rho}$.

In most cases, however, the problem of interest involves a conservative force, usually gravity, and a normal (constraint) force. In these cases the problem is conservative since the normal force does not do work, by definition. One application is to roller coasters and these have recently been discussed in the pedagogical literature, see Pendrill \cite{pendrill} and M{\"u}ller \cite{muller}. In simple mechanical problems given to students the shape of the track is often taken to be circular but in practice this leads to unpleasantly large time variation of the normal force. In practice therefore more complicated shapes are used, in particular the clothoid since the curvature ($1/\rho$) of this curve varies linearly with arc length \cite{muller,miura}. Students that wish to understand the physics behind amusement park experiences in greater detail must therefore be prepared to study curves more advanced than the circle.

Here we will derive the shape of the tracks that produce a constant normal force. Pendrill \cite{pendrill} formulates the problem but does not solve it. Auelmann \cite{auelmann} has studied the effect of adding a normal force of constant magnitude to the central gravitational force of the Kepler problem. He showed that this problem is integrable and derived some general results. Adding a normal force in space requires the firing of rockets perpendicular to the velocity. The much more natural problem of a constant gravitational force, a good approximation in amusement parks, with the normal force coming from a smooth track, also turns out to be integrable, as we will show in this article.

\section{The problem and its conserved quantities}
 Gonz{\'a}lez-Villaneuva \etal\ \cite{gonzalez} found that the trajectories of the Kepler problem are arcs of circles in velocity space. As it turns out the trajectories of our problem are circles in acceleration space. Integrating once we thus find that the trajectories in velocity space correspond to the conic section solutions of the Kepler problem. The shape of the tracks that lead to a normal force of constant magnitude $N$ are then obtained by one further integration.

We write the equation of motion,
\begin{equation}
\label{eq.of.mot.vec}
  m\dot{\vecv}=-mg\,\vece_y+N\,\vece_\theta ,
\end{equation}
where $\vece_\theta = -\sin\theta\,\vece_x + \cos\theta\,\vece_y$, and $\theta$ is the angle between the velocity vector and the $x$-axis. We use polar coordinates in velocity space and put $\vecv = \dot x\,\vece_x + \dot y\,\vece_y = \sqrt{{\dot x}^2 + {\dot y}^2}( \cos\theta\,\vece_x + \sin\theta\,\vece_y) =r \,\vece_r$. Clearly the vector $\vece_\theta$ is perpendicular to the velocity $r\,\vece_r$ and thus normal to the trajectory.

We note that (\ref{eq.of.mot.vec}) means that the tip of $m\dot{\vecv}$ is on a circle of radius $N$ centered on the tip of the gravitational force vector $-mg\,\vece_y$. The two Cartesian components of this vector equation are,
\begin{equation}
  m\ddot{x}=-N\dot{y}/\sqrt{\dot{x}^2+\dot{y}^2},\quad
  m\ddot{y}=-mg+N\dot{x}/\sqrt{\dot{x}^2+\dot{y}^2} .
\end{equation}
We introduce the following dimensionless variables,
\begin{equation}
\label{eq.dim.less.var}
  x\leftarrow\frac{g}{v_0^2}x,\quad y\leftarrow\frac{g}{v_0^2}y,\quad \dfd t\leftarrow\frac{g}{v_0}\dfd t,\quad
  \vx\leftarrow\frac{\dot{x}}{v_0},\quad
  \vy\leftarrow\frac{\dot{y}}{v_0} ,
\end{equation}
and put,
\begin{equation}
  \lambda=N/mg .
\end{equation}
Using this the equations of motion are,
\begin{equation}
\label{eq.of.mot.dimless.xy}
  \dot{v}_x=-\lambda\vy/\sqrt{\vx^2+\vy^2},\quad
  \dot{v}_y=-1+\lambda\vx/\sqrt{\vx^2+\vy^2}.
\end{equation}
We denote the dimensionless speed by $r$ so that,
\begin{equation}
  \vx=r\cos{\theta},\quad\vy=r\sin{\theta}.
\end{equation}
Transforming Eqs.\ (\ref{eq.of.mot.dimless.xy}) to velocity space polar coordinates $r, \theta$ gives,
\begin{equation}\label{eq.of.mot.dimless.rtheta}
\dot r = -\sin\theta,\quad
r \dot \theta = \lambda - \cos\theta ,
\end{equation}
which means that,
\begin{equation}
  \dfd r=-\sin\theta\,\dfd t,\quad r\,\dfd\theta=\left(\lambda-\cos\theta\right)\dfd t
\end{equation}
Add the first of these multiplied with $\lambda-\cos\theta$ to the second multiplied with $\sin\theta$ to get,
\begin{equation}
\left(\lambda-\cos\theta\right)  \dfd r + \sin\theta \,r\, \dfd\theta=0 \; \Rightarrow \;
  \dfd\left[ r\left(\lambda-\cos\theta\right)\right]=0 .
\end{equation}
Using the second of the Eqs.\ (\ref{eq.of.mot.dimless.rtheta}) then shows that,
\begin{equation}
\dfd\left[r\left(\lambda-\cos\theta\right)\right]=\dfd(r^2\dot{\theta})=0 .
\end{equation}
Evidently,
\begin{equation}\label{eq.conserv.ang.mom.vel}
r\left(\lambda-\cos\theta\right) = r^2\dot{\theta} = L ,
\end{equation}
is a conserved quantity (constant of the motion).
One recognizes this as a quantity that corresponds to conserved angular momentum (or sectorial velocity) in velocity space.

Obviously there is another conserved quantity for this problem: the energy
\begin{equation}
\label{eq.energy.dim.less}
  \frac{r^2}{2}+y = E .
\end{equation}
Without loss of generality we can assume initial conditions such that $E=0$. We then have,
\begin{equation}
\label{eq.y.r.rel}
y=-\frac{r^2}{2}.
\end{equation}
We thus have a two dimensional problem with two conserved quantities and this means that we have an integrable problem.

\section{Velocity space trajectories}
From Eq.\ (\ref{eq.conserv.ang.mom.vel}) we find that,
\begin{equation}\label{eq.con.sec.vel.space}
r = \frac{L/\lambda}{1 - (1/\lambda)\cos\theta},
\end{equation}
$r$ and $\theta$ being polar coordinates for the velocity vector. This is easily recognized as the expression for a conic section (see \eg\ Goldstein \cite{BKgoldstein}) where $1/\lambda$ corresponds to the eccentricity, and where the origin corresponds to one of the foci. We have arrived at the conclusion that the velocity space trajectory, the hodograph, of a constant normal force trajectory is a conic section, \ie\ an ellipse, a parabola, or a hyperbola.

We note that, only in the elliptic case, when the eccentricity $1/\lambda < 1$ will the speed, $r= \dot s$, be finite for all angles. For $\lambda = 1$, the parabolic case, the speed goes to infinity for $\cos\theta \rightarrow 1$. When $\lambda < 1$ the speed goes to infinity along the asymptotes $\theta \rightarrow \pm \arccos\lambda$.

\section{Integrating the problem}
Since this two dimensional problem has two constants of the motion it is fully integrable. Here we use this to calculate the trajectories for the various cases that arise. We chose initial conditions so that, at time $t=0$, we have, $r_0=1$, and $x_0=0$. According to (\ref{eq.y.r.rel}) we then also have $y_0= -r_0^2/2=-1/2$.

\subsection{The trivial straight line case}
The simplest case is the trivial case of $L=0$, or $\lambda-\cos\theta_0=0$.
From (\ref{eq.conserv.ang.mom.vel}) we then have that $\dot\theta=0$ so $\theta$ is constant $=\theta_0$. Then $\dot r = -\sin\theta_0$ and we have,
\begin{equation}
  \theta=\theta_0,\quad
 r=1-\sin\theta_0 t.
\end{equation}
Using that $r=\dot s$, where $s$ is arc length, we find,
\begin{equation}
 s(t) =t-\sin\theta_0\, \frac{t^2}{2},\quad \;
 x(t) =\cos\theta_0 \left( t-\sin\theta_0\, \frac{t^2}{2} \right),\quad
 y (t) =-\frac{(1-\sin\theta_0\, t)^2}{2}.
\end{equation}
The second of these follow from time integration of $v_x=\dot x = r \cos\theta_0$, and the second from (\ref{eq.y.r.rel}). These results can also be written,
\begin{equation}
 x(s) =\cos\theta_0 \,s,\quad
 y(s) =-1/2 + \sin\theta_0 \, s,\quad \;
 y(x) = \tan\theta_0 \, x -1/2 ,
\end{equation}
so we are dealing with a straight line of slope $\tan\theta_0 = \sqrt{1-\lambda^2}/\lambda$.

\subsection{The non-trivial cases}
If $L=\lambda-\cos\theta_0 \neq0$, then
\begin{equation}
\label{eq.r.as.qoutient.init.cond}
  r=\frac{\lambda-\cos\theta_0}{\lambda-\cos\theta},
\end{equation}
according to Eq.\ (\ref{eq.conserv.ang.mom.vel}). Using this and the second of Eqs.\ (\ref{eq.of.mot.dimless.rtheta}) one finds that $r\, \dfd \theta = (\lambda - \cos\theta) \dfd t$ and hence that,
\begin{equation}
\label{eqs.dtsx}
  \dfd t=\frac{\lambda-\cos\theta_0}{(\lambda-\cos\theta)^2}\dfd\theta,\quad
  \dfd s=\frac{(\lambda-\cos\theta_0)^2}{(\lambda-\cos\theta)^3}\dfd\theta,\quad
  \dfd x=\frac{(\lambda-\cos\theta_0)^2\cos\theta}{(\lambda-\cos\theta)^3}\dfd\theta.
\end{equation}
The  second of these is obtained by noting that $\dfd s / \dfd t = r$, and the third follows from $\dfd x/\dfd t = r\, \cos\theta$.

Finding the functions $t(\theta,\lambda), s(\theta,\lambda), x(\theta,\lambda)$ thus requires the following integrals to be done:
\begin{equation}
\label{eq.t.integral}
  t=(\lambda-\cos\theta_0)\int_{\theta_0}^\theta\frac{\dfd \theta^\prime}{(\lambda-\cos\theta^\prime)^2}
\end{equation}
\begin{equation}
\label{eq.s.integral}
  s=(\lambda-\cos\theta_0)^2\int_{\theta_0}^\theta\frac{\dfd \theta^\prime}{(\lambda-\cos\theta^\prime)^3}
\end{equation}
\begin{equation}
\label{eq.x.integral}
  x=(\lambda-\cos\theta_0)^2\int_{\theta_0}^\theta\frac{\cos\theta^\prime\dfd \theta^\prime}{(\lambda-\cos\theta^\prime)^3}.
\end{equation}
A complete expression for the trajectory is obtained by noting that the function $y(\theta,\lambda)$  is given by,
\begin{equation}
\label{eq.y.integral}
  y=(\lambda-\cos\theta_0)^2\frac{-1}{2(\lambda-\cos\theta)^2},
\end{equation}
according to Eqs.\ (\ref{eq.y.r.rel}) and (\ref{eq.r.as.qoutient.init.cond}).

When doing the integrals (\ref{eq.t.integral}) - (\ref{eq.x.integral}) it is convenient to distinguish
the following cases:
\begin{eqnarray}
\label{case1}
 \mbox{Case 1:} & \quad\lambda>1,\quad-\pi<\theta<\pi,\quad & \theta_0 = 0\\
 \mbox{Case 2:} & \quad\lambda>1,\quad0<\theta<2\pi,\quad & \theta_0 = \pi\\
 \mbox{Case 3:} & \quad\lambda=1,\quad0<\theta<2\pi,\quad & \theta_0 = \pi\\
 \mbox{Case 4:} & \quad 0\le\lambda<1,\quad
  -\arccos\lambda<\theta<\arccos\lambda, & \quad \theta_0 = 0\\
\label{case5}
 \mbox{Case 5:} & \quad 0\le\lambda<1,\quad
  \arccos\lambda<\theta<2\pi-\arccos\lambda, & \quad \theta_0 = \pi
\end{eqnarray}
The explicit results are given in Appendix A.

\subsection{The elliptic case}
It turns out that the complicated expressions obtained by direct integration using the variable $\theta$ become much simpler if one introduces a suitably chosen new integration variable. In the Kepler problem this new variable is called the eccentric anomaly \cite{BKgoldstein,BKroy}.

For Case 1 we chose,
\begin{equation}
\psi=2\arctan\left(
    \sqrt{\frac{\lambda+1}{\lambda-1}}\frac{\sin\theta}{1+\cos\theta}
    \right),
\end{equation}
so that, $-\pi <\psi< \pi$, is in the same range as $\theta$.
One then finds that \eg\ $\dfd t$, as given in Eqs.\ (\ref{eqs.dtsx}), becomes,
\begin{equation}\label{eq.dt.eccentric}
\dfd t = (\lambda-\cos\theta_0) \frac{(\lambda-\cos\psi)}{(\lambda^2 -1)^{3/2}}\dfd\psi .
\end{equation}
Also the other integrals simplify considerably.

For Case 1 we then get the following explicit results for the trajectory as a function of the parameter $\psi$,
\begin{eqnarray}
\label{eq.t1}
  t(\psi,\lambda)&=&\frac{(\lambda-1)}{(\lambda^2-1)^{3/2}}
    \left(\lambda\psi+\sin\psi\right) , \\
  s(\psi,\lambda)&=&\frac{(\lambda-1)^2}{2(\lambda^2-1)^{5/2}}
  \left[(2\lambda^2+1)\psi+(4\lambda+\cos\psi)\sin\psi\right] , \\
  x(\psi,\lambda)&=&\frac{(\lambda-1)^2}{2(\lambda^2-1)^{5/2}}
  \left\{ 3\lambda\psi+\left[2(\lambda^2+1)+\lambda\cos\psi\right]\sin\psi \right\} , \\
  y(\psi,\lambda)&=&-\frac{(\lambda-1)^2}{2(\lambda^2-1)^2}
  \left(\lambda+\cos\psi\right)^2 ,
\end{eqnarray}
from Eqs.\ (\ref{eq.t.integral}) - (\ref{eq.y.integral}). One notes that one can allow $\psi$ to take values from $-\infty$ to $\infty$ here.
\begin{figure}
\centering
\includegraphics[width=350pt]{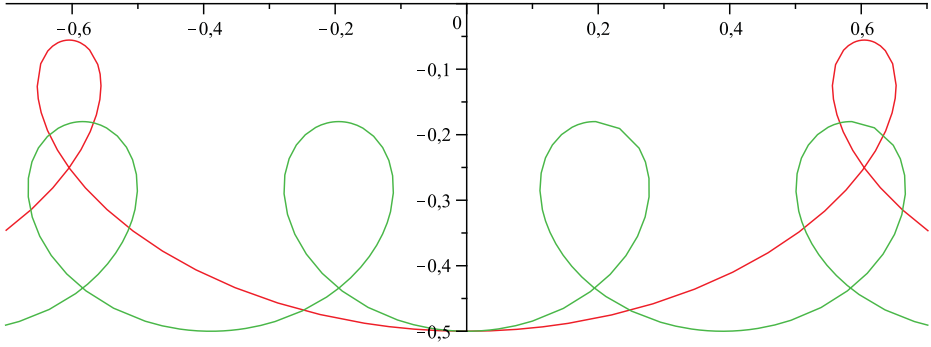}
\vspace{1ex} \caption {\label{fig.TrajCase1} Plot of Case 1 trajectories for $\lambda=N/mg=2$ (red curve with two loops) and for $\lambda=4$ (green curve with four loops). As in all our plots the $x$-axis is horizontal and the $y$-axis vertically upwards. }
\end{figure}

Two such curves are plotted in Fig.\ \ref{fig.TrajCase1}. One notes the limit $\lambda \rightarrow 1+$ which corresponds to a straight horizontal line. For this case therefore $x(\psi,1+)=s(\psi,1+)$. In the limit $\lambda\rightarrow\infty$ the curve approaches a circle.

Let us get some explicit numbers out. Using the factors given in Eq.\ (\ref{eq.dim.less.var}), $v_0$ being the speed at the bottom of the loop, we now find that the time $T$ that it takes to traverse a complete period is given by,
\begin{equation}\label{eq.period}
T = \frac{v_0}{g}[ t(2\pi,\lambda) - t(0,\lambda)] = \frac{v_0}{g} \frac{2\lambda\pi}{\sqrt{\lambda^2-1}(\lambda+1)}
\end{equation}
where the function $t(\psi,\lambda)$ is given by Eq.\ (\ref{eq.t1}) above. Similarly the total length $L$ of one period of the track is given by,
\begin{equation}\label{eq.length}
L = \frac{v_0^2}{g}[ s(2\pi,\lambda) - s(0,\lambda)] = \frac{v_0^2}{g} \frac{(2\lambda^2+1)\pi}{\sqrt{\lambda^2-1}(\lambda+1)^2},
\end{equation}
the horizontal width $W$ is,
\begin{equation}\label{eq.width}
W = \frac{v_0^2}{g}[ x(2\pi,\lambda) - x(0,\lambda)] = \frac{v_0^2}{g} \frac{3\lambda\pi}{\sqrt{\lambda^2-1}(\lambda+1)^2},
\end{equation}
and the vertical height $H$ is,
\begin{equation}\label{eq.height}
H = \frac{v_0^2}{g}[ y(\pi,\lambda) - y(0,\lambda)] = \frac{v_0^2}{g} \frac{2\lambda}{(\lambda+1)^2}.
\end{equation}
We also find from (\ref{eq.r.as.qoutient.init.cond}) that $v(\theta) = v_0 (\lambda -\cos\theta_0)/(\lambda-\cos\theta)$, so the speed $v_{\pi}$ at the top of the loop ($\theta=\pi$) is given by,
\begin{equation}\label{eq.speed}
v_{\pi} =v_0  \frac{(\lambda-1)}{(\lambda+1)}.
\end{equation}
Let us take the speed at the bottom of the track to be $v_0 = 20\,$m/s (this corresponds to 72 km/h) and use $g=9.81\,$m/s$^2$. One then finds that, for $\lambda=2$, \ie\ normal force $N=2mg$, $T=4.93\,$s, $L=73.96\,$m, $W=49.30\,$m, $H=18.12\,$m, and $v_{\pi}= 6.67\,$m/s. For $\lambda=4$ this becomes $T=2.65\,$s, $L=43.66\,$m, $W=15.88\,$m, $H=13.05\,$m, and $v_{\pi}= 12.00\,$m/s.

Finally we note that Case 2, the other elliptic case with $\lambda>1$, simply gives similar displaced trajectories compared to Case 1. For Case 2 the top of a loop will be at, $x=0, y=-1/2$, as in Figs.\ \ref{fig.TrajCase3} and \ref{fig.TrajCase5} below. We therefore proceed directly to the remaining cases.

\subsection{The other cases}
For all the cases with $\lambda \le 1$ the trajectory in velocity space goes to infinity. These tracks can therefore not constitute a real roller coaster with constant normal force in themselves. However as parts of more complicated tracks, spliced together with clothoids, they might be of interest.

\begin{figure}
\centering
\includegraphics[width=200pt]{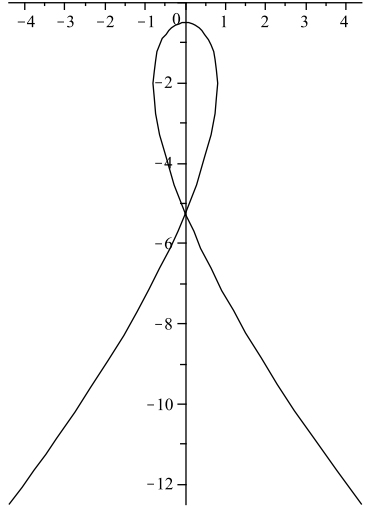}
\vspace{1ex} \caption {\label{fig.TrajCase3} Plot of the unique Case 3 trajectory with normal force equal to gravity ($\lambda=1$). }
\end{figure}

For Case 3, the parabolic case, we have $\lambda=1$ and $\theta_0=\pi$, so that $(\lambda-\cos\theta_0)=2$. The variable that simplifies the integrations turns out to be,
\begin{equation}
  \omega=-\frac{\sin\theta}{1-\cos\theta},
\end{equation}
with range $-\infty<\omega<\infty$.
From Eqs.\ (\ref{eq.t.integral}) - (\ref{eq.y.integral}) one finds,
\begin{eqnarray}
  t(\omega)&=&\frac{1}{3}
    (\omega^2+3)\omega , \\
  s(\omega)&=&\frac{1}{15}
    (3\omega^4+10\omega^2+15)\omega , \\
  x(\omega)&=&\frac{1}{5}
    (\omega^4-5)\omega , \\
  y(\omega)&=&-\frac{1}{2}
    (\omega^2+1)^2 ,
\end{eqnarray}
for this track, in terms of the parameter $\omega$. Fig.\ \ref{fig.TrajCase3} shows the shape of this one-loop track.

\begin{figure}
\centering
\includegraphics[width=200pt]{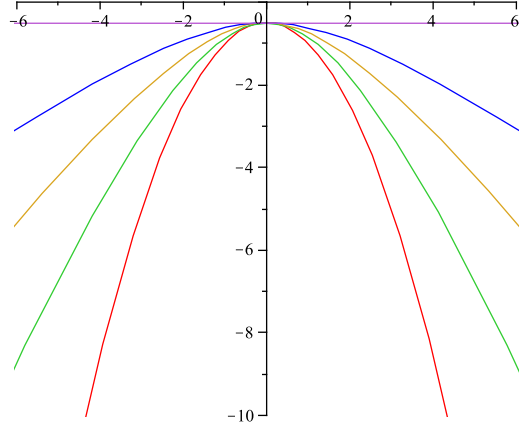}
\vspace{1ex} \caption {\label{fig.TrajCase4} Plot of a Case 4 trajectories. The lowest (innermost) red curve is the parabola of free fall corresponding to $\lambda=0$. Above that one the curves correspond to $\lambda=N/mg=1/4$ (green), $\lambda=1/2$ (brown), $\lambda=3/4$ (blue) and finally the horizontal straight line (violet) corresponds to $\lambda\rightarrow 1$.}
\end{figure}

We now come to the two hyperbolic cases. For Case 4 we have $\lambda<1$ and $\theta_0=0$ so $(\lambda-\cos\theta_0)=(\lambda-1)$. Let,
\begin{equation}
\chi=2\arctanh\left(
    \sqrt{\frac{1+\lambda}{1-\lambda}}\frac{\sin\theta}{1+\cos\theta}
    \right) ,
\end{equation}
with range $-\infty<\chi<\infty$. Then,
\begin{eqnarray}
  t(\chi,\lambda)&=&\frac{(\lambda-1)}{(1-\lambda^2)^{3/2}}
    \left(\lambda\chi+\sinh\chi\right) , \\
  s(\chi,\lambda)&=&-\frac{(\lambda-1)^2}{2(1-\lambda^2)^{5/2}}
  \left[(2\lambda^2+1)\chi+(4\lambda+\cosh\chi)\sinh\chi\right] , \\
  x(\chi,\lambda)&=&-\frac{(\lambda-1)^2}{2(1-\lambda^2)^{5/2}}
  \left\{ 3\lambda\chi+\left[2(\lambda^2+1)+\lambda\cosh\chi\right] \sinh\chi\right\} , \\
  y(\chi,\lambda)&=&-\frac{(\lambda-1)^2}{2(1-\lambda^2)^2}
  \left(\lambda+\cosh\chi\right)^2 .
\end{eqnarray}
The shape of some of these tracks are shown in Fig.\ \ref{fig.TrajCase4}. One notes that in the limiting case of $\lambda \rightarrow 0$ this trajectory becomes,
\begin{eqnarray}\label{eq.parabola.lambda.zero}
t(\chi)=-\sinh\chi,\quad s(\chi)=-(1/2)(\cosh\chi\,\sinh\chi  + \chi),\\ x(\chi)=-\sinh\chi, \quad y(\chi) = -(1/2)\cosh^2\chi ,
\end{eqnarray}
the parabola that is the ballistic trajectory of free fall.

One may also note that the limit $\lambda\rightarrow 1-$ when taken in the $\chi$-dependent parts of $s(\chi,\lambda)$ and $x(\chi,\lambda)$ make these identical: $x(\chi,1-) = s(\chi,1-)$. This means that the curve becomes a straight horizontal line.

\begin{figure}
\centering
\includegraphics[width=250pt]{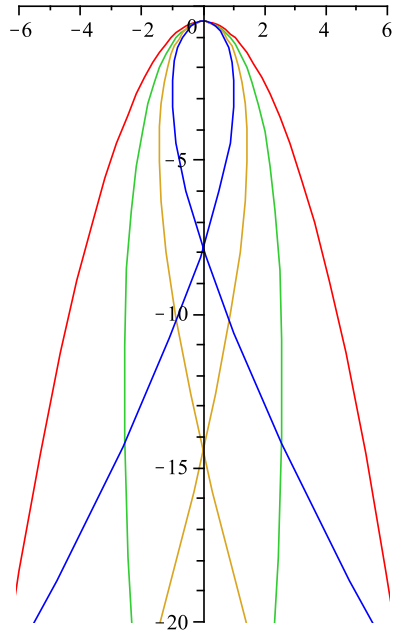}
\vspace{1ex} \caption {\label{fig.TrajCase5} Plot of a Case 5 trajectories. The outermost red curve is the parabola corresponding to $N=\lambda=0$. Inside the parabola there are curves corresponding to the $\lambda$-values $1/4$ (green), $1/2$ (brown), and $3/4$ (blue). The limit $\lambda\rightarrow 1$ is in Fig.\ \ref{fig.TrajCase3}.}
\end{figure}

Finally we have Case 5. Here $\lambda<1$ and $\theta_0=\pi$ so $(\lambda-\cos\theta_0)=(\lambda+1)$.
Let,
\begin{equation}
\eta=-2\arctanh\left(
    \sqrt{\frac{1-\lambda}{1+\lambda}}\frac{\sin\theta}{1-\cos\theta}
    \right),
\end{equation}
with range $-\infty<\eta<\infty$.
Then
\begin{eqnarray}
  t(\eta,\lambda)&=&\frac{(\lambda+1)}{(1-\lambda^2)^{3/2}}
    \left(-\lambda\eta+\sinh\eta\right) , \\
  s(\eta,\lambda)&=&\frac{(\lambda+1)^2}{2(1-\lambda^2)^{5/2}}
  \left[(2\lambda^2+1)\eta-(4\lambda-\cosh\eta)\sinh\eta\right] , \\
  x(\eta,\lambda)&=&\frac{(\lambda+1)^2}{2(1-\lambda^2)^{5/2}}
  \left\{ 3\lambda\eta-\right[ 2(\lambda^2+1)-\lambda\cosh\eta\left] \sinh\eta \right\} , \\
  y(\eta,\lambda)&=&-\frac{(\lambda+1)^2}{2(1-\lambda^2)^2}
  \left(\lambda-\cosh\eta\right)^2 .
\end{eqnarray}
For $\lambda=0$ this curve also becomes a the ballistic parabola. For positive values of $\lambda$ 
this trajectory has a loop and it is shown in Fig.\ \ref{fig.TrajCase5} for some $\lambda$-values.

\section{Conclusions}
We have presented explicit solutions for the problem of which smooth tracks produce a normal force of constant magnitude on a particle that slides along the track under the influence of a constant gravitational field in a vertical plane. The obvious application is to the design of roller coasters. As we hope to have shown, the problem in itself has many interesting and surprising features. One of them is the relation to the Kepler problem of planetary motion. Students can learn various classical techniques of analysis by studying this problem. One notes that a direct numerical approach cannot reveal the fact that there are five (or four, depending on the point of view) qualitatively different regions of parameter values, nor the qualitative features of the corresponding solutions and their limiting behavior.

\appendix
\section{Appendix}
Explicit results for the integrals in Eqs.\ (\ref{eq.t.integral}) - (\ref{eq.x.integral}), without the constants factors $(\lambda-\cos\theta_0)^k$, with $k=1$ or 2, in front, \ie
\begin{equation}\nonumber
  t_i(\theta,\lambda)=\int_{\theta_0}^\theta\frac{\dfd \theta^\prime}{(\lambda-\cos\theta^\prime)^2} ,
\end{equation}
\begin{equation}\nonumber
  s_i(\theta,\lambda)=\int_{\theta_0}^\theta\frac{\dfd \theta^\prime}{(\lambda-\cos\theta^\prime)^3} ,
\end{equation}
\begin{equation}\nonumber
  x_i(\theta,\lambda)=\int_{\theta_0}^\theta\frac{\cos\theta^\prime\dfd \theta^\prime}{(\lambda-\cos\theta^\prime)^3} ,
\end{equation}
are given here. For each of the three integrals there are five cases (\ref{case1})-(\ref{case5}) and
$i=1,2,3,4,5$ represents the Case number.

One finds:
\begin{eqnarray}\nonumber
  t_1(\theta,\lambda)&=&\frac{1}{(\lambda^2-1)}\left[
    \frac{2\lambda}{\sqrt{\lambda^2-1}}\arctan\left(
    \sqrt{\frac{\lambda+1}{\lambda-1}}\frac{\sin\theta}{1+\cos\theta}
    \right)+\frac{\sin\theta}
    {\left(\lambda-\cos\theta\right)}\right]\\
    \nonumber
  t_2(\theta,\lambda)&=&\frac{1}{(\lambda^2-1)}\left[
    -\frac{2\lambda}{\sqrt{\lambda^2-1}}\arctan\left(
    \sqrt{\frac{\lambda-1}{\lambda+1}}\frac{\sin\theta}{1-\cos\theta}
    \right)+\frac{\sin\theta}
    {\left(\lambda-\cos\theta\right)}\right]\\
    \nonumber
  t_3(\theta)&=&-\frac{1}{3}
  \frac{\left(2-\cos\theta\right)\sin\theta}
       {\left(1-\cos\theta\right)^2}\\
       \nonumber
  t_4(\theta,\lambda)&=&\frac{1}{(\lambda^2-1)}\left[
    -\frac{2\lambda}{\sqrt{1-\lambda^2}}\arctanh\left(
    \sqrt{\frac{1+\lambda}{1-\lambda}}\frac{\sin\theta}{1+\cos\theta}
    \right)+\frac{\sin\theta}
    {\left(\lambda-\cos\theta\right)}\right]\\
    \nonumber
  t_5(\theta,\lambda)&=&\frac{1}{(\lambda^2-1)}\left[
    -\frac{2\lambda}{\sqrt{1-\lambda^2}}\arctanh\left(
    \sqrt{\frac{1-\lambda}{1+\lambda}}\frac{\sin\theta}{1-\cos\theta}
    \right)+\frac{\sin\theta}
    {\left(\lambda-\cos\theta\right)}\right]
\end{eqnarray}
\begin{eqnarray}
\nonumber
  s_1(\theta,\lambda)&=&\frac{1}{(\lambda^2-1)^2}\left[
    \frac{2\lambda^2+1}{\sqrt{\lambda^2-1}}\arctan\left(
    \sqrt{\frac{\lambda+1}{\lambda-1}}\frac{\sin\theta}{1+\cos\theta}
    \right)+\frac{\left(4\lambda^2-3\lambda\cos\theta-1\right)\sin\theta}
    {2\left(\lambda-\cos\theta\right)^2}\right]\\
    \nonumber
  s_2(\theta,\lambda)&=&\frac{1}{(\lambda^2-1)^2}\left[
    -\frac{2\lambda^2+1}{\sqrt{\lambda^2-1}}\arctan\left(
    \sqrt{\frac{\lambda-1}{\lambda+1}}\frac{\sin\theta}{1-\cos\theta}
    \right)+\frac{\left(4\lambda^2-3\lambda\cos\theta-1\right)\sin\theta}
    {2\left(\lambda-\cos\theta\right)^2}\right]\\
    \nonumber
  s_3(\theta)&=&-\frac{1}{15}
  \frac{\left(2\cos\theta^2-6\cos\theta+7\right)\sin\theta}
       {\left(1-\cos\theta\right)^3}\\
       \nonumber
  s_4(\theta,\lambda)&=&\frac{1}{(\lambda^2-1)^2}\left[
    -\frac{2\lambda^2+1}{\sqrt{1-\lambda^2}}\arctanh\left(
    \sqrt{\frac{1+\lambda}{1-\lambda}}\frac{\sin\theta}{1+\cos\theta}
    \right)+\frac{\left(4\lambda^2-3\lambda\cos\theta-1\right)\sin\theta}
    {2\left(\lambda-\cos\theta\right)^2}\right]\\
    \nonumber
  s_5(\theta,\lambda)&=&\frac{1}{(\lambda^2-1)^2}\left[
    -\frac{2\lambda^2+1}{\sqrt{1-\lambda^2}}\arctanh\left(
    \sqrt{\frac{1-\lambda}{1+\lambda}}\frac{\sin\theta}{1-\cos\theta}
    \right)+\frac{\left(4\lambda^2-3\lambda\cos\theta-1\right)\sin\theta}
    {2\left(\lambda-\cos\theta\right)^2}\right]
\end{eqnarray}
\begin{eqnarray}
\nonumber
  x_1(\theta,\lambda)&=&\frac{1}{(\lambda^2-1)^2}\left[
    \frac{3\lambda}{\sqrt{\lambda^2-1}}\arctan\left(
    \sqrt{\frac{\lambda+1}{\lambda-1}}\frac{\sin\theta}{1+\cos\theta}
    \right)+\frac{\left((2\lambda^2+1)\lambda-
      (\lambda^2+2)\cos\theta\right)\sin\theta}
    {2\left(\lambda-\cos\theta\right)^2}\right]\\
    \nonumber
  x_2(\theta,\lambda)&=&\frac{1}{(\lambda^2-1)^2}\left[
    -\frac{3\lambda}{\sqrt{\lambda^2-1}}\arctan\left(
    \sqrt{\frac{\lambda-1}{\lambda+1}}\frac{\sin\theta}{1-\cos\theta}
    \right)+\frac{\left((2\lambda^2+1)\lambda-
      (\lambda^2+2)\cos\theta\right)\sin\theta}
    {2\left(\lambda-\cos\theta\right)^2}\right]\\
    \nonumber
  x_3(\theta)&=&\frac{1}{5}
  \frac{\left(\cos\theta^2-3\cos\theta+1\right)\sin\theta}
       {\left(1-\cos\theta\right)^3}\\
       \nonumber
  x_4(\theta,\lambda)&=&\frac{1}{(\lambda^2-1)^2}\left[
    -\frac{3\lambda}{\sqrt{1-\lambda^2}}\arctanh\left(
    \sqrt{\frac{1+\lambda}{1-\lambda}}\frac{\sin\theta}{1+\cos\theta}
    \right)+\frac{\left((2\lambda^2+1)\lambda-
      (\lambda^2+2)\cos\theta\right)\sin\theta}
    {2\left(\lambda-\cos\theta\right)^2}\right]\\
    \nonumber
  x_5(\theta,\lambda)&=&\frac{1}{(\lambda^2-1)^2}\left[
    -\frac{3\lambda}{\sqrt{1-\lambda^2}}\arctanh\left(
    \sqrt{\frac{1-\lambda}{1+\lambda}}\frac{\sin\theta}{1-\cos\theta}
    \right)+\frac{\left((2\lambda^2+1)\lambda-
      (\lambda^2+2)\cos\theta\right)\sin\theta}
    {2\left(\lambda-\cos\theta\right)^2}\right]
\end{eqnarray}
These complicated expressions simplify considerably after transformation to new integration variables corresponding to eccentric anomalies as discussed above.

\bibliographystyle{unsrt}

\end{document}